\documentstyle[epsfig,12pt]{article}

\newcommand{\nc}{\newcommand}
\nc{\beq}{\begin{equation}} \nc{\eeq}{\end{equation}}
\nc{\beqa}{\begin{eqnarray}} \nc{\eeqa}{\end{eqnarray}}
\def\DS {D\!\!\!\!/}

\def\lsim{\mathrel{\rlap{\lower4pt\hbox{\hskip1pt$\sim$}}
    \raise1pt\hbox{$<$}}}       
\def\gsim{\mathrel{\rlap{\lower4pt\hbox{\hskip1pt$\sim$}}
    \raise1pt\hbox{$>$}}}       

\begin{document}

\title{{\large {\bf Quintessence and Thermal Matter}}}

\author{
S. Hsu\thanks{hsu@duende.uoregon.edu} ,
B. Murray\thanks{bmurray1@darkwing.uoregon.edu} \\
Department of Physics \\
University of Oregon, Eugene OR 97403-5203 \\}
\maketitle

\begin{abstract}
We investigate the effects of thermal interactions on tracking
models of quintessence. We show that even Planck-suppressed
interactions between matter and the quintessence field can alter
its evolution qualitatively. The dark energy equation of state is
in many cases strongly affected by matter couplings. We obtain a
bound on the coupling between quintessence and relativistic relic
particles such as the photon or neutrino.
\end{abstract}

\newpage

\section{Introduction}

Recent evidence \cite{observation} suggests that a large fraction
of the energy density of the universe has negative pressure, or
equation of state with $ w \equiv p /\rho < 0$. One candidate
source of this dark energy is a slowly varying and spatially
homogeneous scalar field called quintessence \cite{darkenergy}.
Because the dark energy redshifts more slowly than ordinary matter
or radiation, it appears that the ratio of energy density of
quintessence to that in ordinary particles must be fine tuned to a
specific infinitesimal value in the early universe in order to
explain its current observed value. One class of models that
ameliorate this problem describe {\it tracker fields}
\cite{tracker} whose evolution is largely insensitive to
initial conditions and at late times begin to dominate the
energy density of the universe with a negative equation of state.
Tracker models have difficulty producing
$w_\phi$ consistent with observational data: they generally imply
$w^{\rm eff}_\phi \gsim -0.7$ , whereas WMAP implies
$w_{\phi}^{\rm eff} < -0.78$ (95\% CL) \cite{observation}.
(Here, effective means as measured observationally, so integrating over
redshifts less than of order $10^3$.)
Nevertheless, they provide an interesting class of models describing
dark energy as a slowly evolving scalar field.
An alternative class of models, which avoids
the extremely flat potentials required at late times of tracker
models, utilizes nonlinear field oscillations that exhibit $w < 0$
\cite{nonlinear}.

Tracker models generally require only a single adjustable parameter. Once
this parameter is appropriately chosen, a wide range of initial
values of the tracker field, $\phi$, and its derivative, $\dot{\phi}$,
result in similar values of its energy density today. This is due to
an attractor-like property of the tracker equations of motion.

In this paper we investigate the effects of interactions between
the quintessence field, $\phi$, and ordinary matter particles in
the early universe. It is important to note that while the zero
temperature potential may be fine tuned in order for the evolution
of $\phi$ to have the attractive properties mentioned above, the
same may not be done with the finite temperature effective
potential. That is to say, once the form of the renormalized zero
temperature potential is determined, no additional freedom remains
to fine-tune away unwanted thermal effects. Therefore, such
effects must be considered. We expect thermal interactions to be
at least of gravitational strength (even in the case where $\phi$
is a ``hidden sector'' field). At minimum, quantum gravity is
likely to produce interactions of the type \cite{Planck} \beq
\label{int} {\beta_i \over M_P} \, \phi \, {\cal L}_i ~~,\eeq
where $M_P$ is the Planck scale, and ${\cal L}_i$ are terms in the
standard model lagrangian, including for example
$$
F_{\mu \nu}^2~,~F_{\mu \nu} \tilde{F}^{\mu \nu} ~,~ \bar{\psi} \DS
\, \psi ~,~ \cdots
$$
where $F$ is the field strength of any gauge field (including the
photon, but not excluding gluons or the W or Z) and $\psi$ is any
fermion field from neutrinos to the top quark. Even if $\phi$ were
a pseudo-Goldstone boson \cite{constraints}, it would be
surprising not to find at least Planck-suppressed violations of
the resulting $~\phi \rightarrow \phi + {\rm constant}~$ symmetry.
String theory, for example, is believed to not exhibit any exact
global symmetries \cite{strings}. Previous constraints on certain
$\beta_i$ are quite strong, where the coupling is to the photon or
gluon \cite{constraints}. However, some $\beta_i$ could be much
larger, such as when the interaction is with the W, Z or even a
neutrino\footnote{Direct coupling to relic neutrinos has been
considered previously \cite{neutrino}.}.

In the early universe matter particles are in thermal equilibrium,
and the interactions in (\ref{int}) produce a thermal mass for
$\phi$ of the form \beq  \label{mass} \left( {\beta_i \over M_P}
\right)^2 \, \phi^2 \, T^4~~, \eeq where $T$ is the temperature.
If the thermal degree of freedom is massive, the thermal effect
goes to zero exponentially as $e^{- m /T}$ when the temperature
drops below the mass $m$. We note that the thermal effects of
interest here are {\it in addition} to any quantum corrections to
the effective potential resulting from the interactions between
matter and quintessence (see, for example, \cite{qloop}). In
general, the bare parameters of quintessence models must be
fine-tuned in order to obtain potentials of the necessary type. We
assume here that this fine-tuning is achieved (whatever its
consequences for the plausibility of the model) and focus on
thermal effects which must also arise.

Although $\phi$ may not itself be in equilibrium, nevertheless
its dynamical evolution will be affected by these thermal
interactions, just as for the axion field near the QCD phase
transition \cite{axion}. We can derive the correction (\ref{mass})
to the effective potential for $\phi$ as follows. Let the cold
$\phi$ field be a static, external source for a Euclidean path
integral describing the thermal degrees of freedom. The timelike
boundary conditions for the path integral have period given by the
inverse temperature. Performing the integration over the thermal
fields yields a contribution to the effective potential for
$\phi$, and the usual perturbative analysis identifies the leading
effect to be the thermal mass term in (\ref{mass}). In this
calculation, we need never assume that $\phi$ itself is in thermal
equilibrium, yet its effective potential receives
temperature-dependent contributions.

In what follows we will examine how the contribution of
(\ref{mass}) to the tracker potential modifies its evolution. We
can make a simple argument for why (\ref{mass}) is non-negligible
at late times. At late times, the quintessence field must have a
very small mass: $V''(\phi)^{1/2} \lsim H_0 \sim 10^{-33}$ eV, and
contribute of order closure density to $\Omega$: $V(\phi) \sim
(10^{-3} {\rm eV})^4$, which implies that $\phi \sim M_P$. This
means that the mass term in (\ref{mass}) can be roughly the same
size as $V(\phi)$, up to powers of $\beta_i$.

At early times, (\ref{mass}) also affects the evolution in many
cases. Suppose the tracker potential is given by $V(\phi) =
M^{l+4} \phi^{-l}$, where $l > 4$. Then $V(\phi_*)$ and
(\ref{mass}) are comparable at the minimum of the combined
potential: \beq \phi_* \sim M \left( {M^2 M_P^2 \over \beta^2 T^4}
\right)^{1 \over l+2} ~~,\eeq where the potential energy density
is roughly \beq V (\phi_*) \equiv V_* \sim M^4 \left( { \beta^2
T^4 \over M^2 M_P^2} \right)^{l \over l+2} ~~.\eeq In many cases,
$\phi$ oscillates about the temperature-dependent minimum
$\phi_*$. The oscillation energy redshifts faster than the
potential energy at the minimum, $V_* \sim T^{4l / (l+2)}$, so
$\phi$ simply tracks $\phi_*$ with oscillations that decrease in
amplitude over time. Interestingly, $V_*$ redshifts exactly as the
energy density of the tracker solution
\cite{pr1988} (assuming radiation domination; during a matter
dominated epoch $V_*$ redshifts somewhat faster than the usual
tracker energy density). This means that thermal effects will keep
$\phi$ and its energy density near their desired values, even
though the physics responsible is very different. When the thermal
term eventually either disappears due to the crossing of a
particle mass threshold, or becomes negligible due to redshift,
$\phi$ will merge back to a tracker solution.

\section{Evolution results}

We assume a spatially flat Robertson-Walker universe, with metric
${\rm d}s^2 = {\rm d}t^2 - a^2(t) {\rm d}{\mathbf x}^2$.
The evolution of a scalar
field minimally coupled to gravity in this spacetime is given by
the Klein-Gordon equation: \beq \label{klein}
\ddot{\phi}+3H\dot{\phi} +V'(\phi)=0 ~~, \eeq where a dot denotes
the derivative with respect to cosmic time, and a prime denotes
the derivative with respect to $\phi$. The evolution of the scale
factor is governed by the Friedmann equation: \beq \label{hubble}
H^2 = {\left( \dot{a} \over a \right)}^2 = {8 \pi \over 3 M_P^2}
\left( \rho_m + \rho_r + \rho_{\phi} \right) ~~,\eeq where if $z$
denotes the redshift, then $\rho_m = \rho_c \Omega_m {\left( 1+z
\right)}^3$, $\rho_r = \rho_c \Omega_r {\left( 1+z \right)}^4$,
and $\rho_{\phi} = {1 \over 2} {\dot{\phi}}^2 + V(\phi)$. Here the
subscript $m$ refers to both baryons and cold dark matter, and
the subscript $r$ refers to both photons and neutrinos.
If the universe is spatially flat, then it will always be the case
that $\Omega_m + \Omega_r + \Omega_{\phi} = 1$. Observational data
\cite{observation} currently favor $\Omega_m \sim 0.3$, $\Omega_r
\sim 10^{-4}$ and $\Omega_{\phi} \sim 0.7$.

Equations (\ref{klein}) and (\ref{hubble}) were integrated
numerically for a wide range of $\phi_i$ and $\dot{\phi}_i$ from
an initial redshift of $z_i = 10^{28}$ (temperature $\sim 10^{16}$
GeV), which might plausibly correspond to the end of inflation.
Motivated by the arguments of the previous section, we took
$V(\phi)$ to be \beq \label{potential} V(\phi) = M^{4+l} \phi^{-l}
+ {\left( \beta \over M_P \right)}^2 \phi^2 T^4 ~~, \eeq where $l
> 4$, $\beta$ is a free parameter, and $M$ is constrained such
that $\Omega_{\phi} \sim 0.7$. For example, for $l = 6$ and $\beta
= 0$, $M \sim 4.7 \times 10^6$ GeV. In this simulation, $T(z) \equiv
\sqrt[4]{\rho_r} = \sqrt[4]{\rho_c \Omega_r} (1+z)$ (we are not
precise about the number of relativistic degrees of freedom).
It is worth
noting that, since the second term in (\ref{potential}) arises due
to a loop effect, it should actually appear in (\ref{potential})
multiplied by a constant
of order $10^{-1}$. In the absence of this factor, the quantity
$\beta$ in (\ref{potential}) differs somewhat from the $\beta_i$
in (\ref{int}). This consideration, however, has no effect on the
qualitative picture described below.

Based on our simulations, we make the following observations.

For $\beta = 0$, tracking occurs for a large range of initial
conditions in $\phi$ and $\dot{\phi}$,
as described in \cite{tracker}. In particular, for $l=6$, if
$\phi$ starts
from rest, any $\phi_i$ in the range $10^{-18}M_P \lsim \phi_i \lsim
10^{-2}M_P$ will be on track by today. In general,
the limits for $\phi_i$ that will be on track by today,
assuming $\dot{\phi}_i = 0$, are found
by solving $\rho_{\phi i} =  M^{l+4} \phi_i^{-l}$ for $\phi_i$,
where $\rho_{\phi i}$ is the initial energy density in $\phi$.
By noting that $M \sim
(\rho_{\phi o} M_P^l)^{1/(4+l)}$, where $\rho_{\phi o}$ is
the present energy density in $\phi$,
it is straightforward to see
that these limits on $\phi_i$
depend on $l$:
\beq \label{phi_i} \phi_i \sim M_P \left(\rho_{\phi o}
\over \rho_{\phi i} \right)^{1 \over l} ~~.\eeq
The minimum value of $\phi_i$ that will be on track by today,
$\phi_{i,min}$, is then simply found by setting $\rho_{\phi i}$
equal to its maximum value. For an initial redshift of $10^{28}$,
this corresponds roughly to $\rho_{\phi_i} \sim 10^{-4} \rho_{B i}$,
where $\rho_{B i}$
is the energy density of dominant background component,
radiation at this redshift ($\rho_r(z=10^{28})\sim10^{61}$ GeV$^4$).
Similarly, the maximum
value of $\phi_i$ that will be on track by today, $\phi_{i,max}$,
is found by setting $\rho_{\phi i}$
equal to its minimum value, which is roughly the background
energy density at equality, $\rho_{eq} \sim 10^{-37}$ GeV$^4$.

For $\beta \not= 0$, $\phi_{i,min}$ is essentially unchanged
because the first term in (\ref{potential}) is dominant for
$\rho_{\phi_i} \sim 10^{-4} \rho_{B i}$, unless $\beta$ is made very
large ($\beta \sim 10^{20}$).
$\beta$ this large will not be discussed further in this paper.
For sufficiently small $\beta$, $\phi_{i,max}$
is also left unchanged. Let $\beta_c \equiv (10^{-4})^{1/2}
(\rho_{eq}/
\rho_{\phi o})^{1/l} \sim 10^{(11-2l)/l}$.
Then for $\beta = \beta_c$ and $\phi_i = \phi_{i,max}$,
$\rho_{\phi i} = 10^{-4} \rho_{B i}$. But
$\rho_{\phi i} \leq 10^{-4} \rho_{B i}$, and $\rho_{\phi}
\propto {\beta}^2{\phi}^2$ for $\phi \gg \phi_*$.
Therefore, for
$\beta \gsim \beta_c$, $\phi_{i,max} \propto 1/\beta$. The net
result is that the range of $\phi_i$ that will be on track by today
(with $\dot{\phi_i}=0$) is independent of $\beta$ for
$\beta \lsim \beta_c$ and goes like $1/\beta$ for
$\beta \gsim \beta_c$.

In addition to affecting the range of $\phi_i$ that track,
the choice of $\beta$ qualitatively affects the dynamics of
$\phi$. Note that $10^{-2} \lsim \beta_c \lsim 1$. In examining
the dynamics of $\phi$ there again seems to be a critical value
of $\beta$. Although this critical value seems to be $\sim 10^{-1}$,
it is not clear whether it is equal to $\beta_c$. As mentioned
above, a factor of order $10^{-1}$ was not included in (\ref{potential}).
Therefore, in what follows reference will be made to $\beta_c$,
which is meant to indicate a $\beta$ in the range
$10^{-2} \lsim \beta \lsim 1$.

\begin{center}
{\scalebox{1}{\includegraphics{plot_rho.epsi}}}
\end{center}
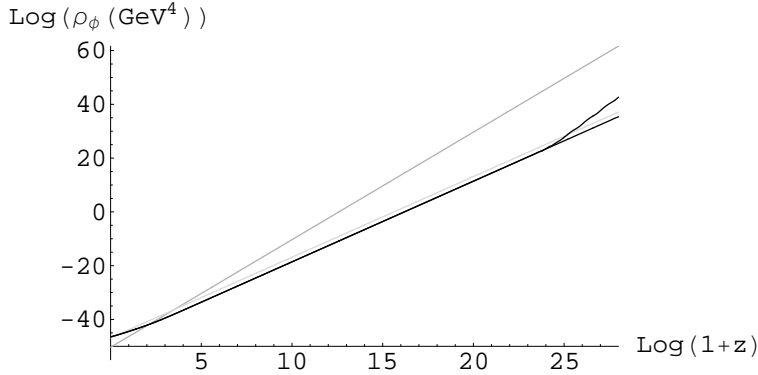
\small FIG. 1. The evolution of $\rho_{\phi}$ is shown for
$\beta=3$, $l=6$ and an intermediate $\rho_{\phi i}$ ($\phi_i \ll
\phi_{* i}$). $\rho_{\phi}$ for $\phi_i = \phi_{* i}$
(corresponding to the tracker solution for $\beta=3$) is also
plotted for reference, as are $\rho_r$ (medium gray) and $\rho_m$
(light gray). Note that $\rho_{\phi}$ cannot decrease below the
the energy density of the tracker solution (it is always larger
than $V_*$) and subsequently freeze as it does for $\beta=0$.
Also, note that, at high redshift, small oscillations can be seen
in $\rho_{\phi}$ corresponding to oscillations of $\phi$ about
$\phi_*$ (as in Figure 2).

\normalsize
\begin{center}
{\scalebox{1}{\includegraphics{plot_rholate.epsi}}}
\end{center}
\small
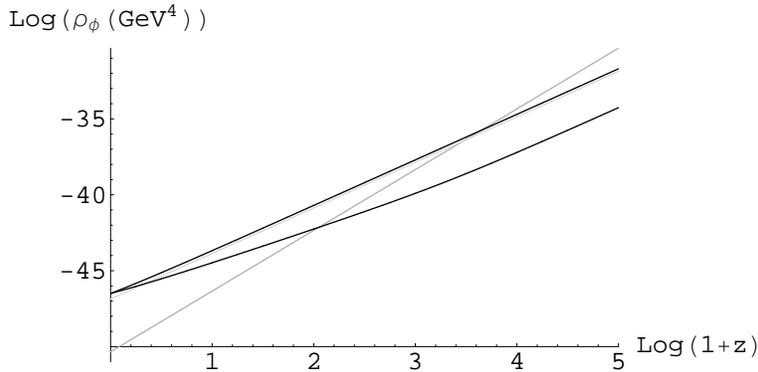
FIG. 2. The late time behavior of the tracker solution
is shown for $\beta=0$
(lower curve) and $\beta=10^2$ (upper curve) with $l=6$.
Again $\rho_r$ (medium gray)
and $\rho_m$ (light gray) are plotted for reference.
Note that while $\rho_{\phi}(\beta=0)$ has a rather shallow
slope today, $\rho_{\phi}(\beta=10^2)$ cannot because
$\rho_{\phi} > V_*$, and $V_*(\beta=10^2) > \rho_m$.
Also, note that the redshift at which $\rho_{\phi}$
begins to dominate depends on $\beta$.
\normalsize

For $\beta \lsim \beta_c$, the behavior of $\phi(z)$ is essentially
just that described in \cite{tracker}, with the one additional
constraint that $\rho_{\phi}(z) > V_*(z)$, at all times
(see Figures 1 and 2).
For $\beta \gsim \beta_c$, $\phi(z)$ oscillates about $\phi_*(z)$
(see Figure 3).
The period of these oscillations in $\phi(z)$ decreases
exponentially with the scale factor, while the amplitude
decreases monotonically, but not exponentially, per se. As either
side of the potential (\ref{potential}) are made steeper;
i.e., either $l$ or $\beta$ are increased, the period of
oscillation decreases, as does the amplitude. The fact that
these oscillations have been damped out by today and that
$V_*$ redshifts at the same rate (for RD) or faster (for MD)
than the tracker solution is what allows for
tracking to occur, even with $\beta \gsim \beta_c$. Adding extra terms
to (\ref{potential}), which cause $V_*$ to redshift slower
than the tracker solution during RD will be discussed below in
Section 3.

\begin{center}
{\scalebox{1}{\includegraphics{plot_phi.epsi}}}
\end{center}
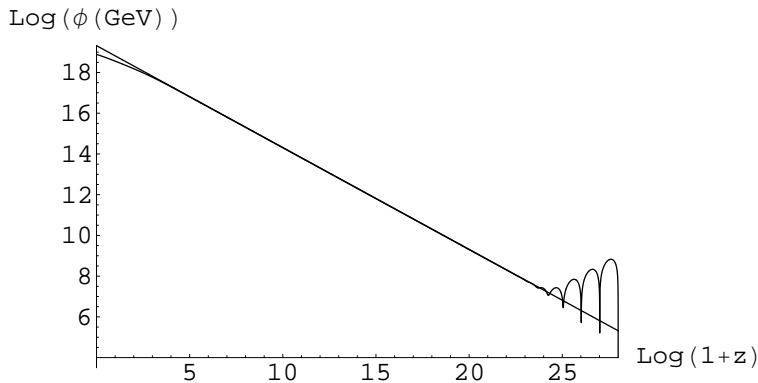
\small FIG. 3. The evolution of $\phi$ is shown for $\beta=3$,
$l=6$ and an intermediate $\rho_{\phi i}$ ($\phi_i \ll \phi_{*
i}$). $\phi_*$ is also plotted for reference. Note that $\phi$
oscillates about $\phi_*$ at large redshift; $\phi$ follows
$\phi_*$ closely for most of its evolution, but for small redshift
it begins to fall behind. \normalsize

For $\beta \lsim \beta_c$, the scale $M$ in (\ref{potential}) is
simply equal to its value for $\beta = 0$; i.e.,
$M \sim (\rho_{\phi o} M_P^l)^{1/(4+l)}$. For $\beta \gsim \beta_c$,
$M$ slowly decreases as $\beta$ is increased, e.g. for
$l=6$ and $\beta = 0$, $M \sim 4.7 \times 10^6$ GeV, whereas
for $l=6$ and $\beta = 10^3$, $M \sim 1.0 \times 10^6$ GeV.

We define the scalar field equation of state: \beq \label{wphi}
w_{\phi} \equiv {1+z \over 3 \rho_{\phi}}{{\rm
d}\rho_{\phi}\over{\rm d}z} -1 ~~. \eeq The definition
(\ref{wphi}) coincides with the usual definition of the scalar
field equation of state: \beq \label{wphiusual} w_{\phi}^{{\rm
usual}} \equiv {p_{\phi} \over \rho_{\phi}}={{{1 \over 2}
\dot{\phi}^2 - V(\phi)}\over {{1 \over 2} \dot{\phi}^2 + V(\phi)}}
~~ \eeq as a consequence of energy conservation. However, because
we did not take into account the back reaction on matter and
radiation of the $\phi$ interaction, there are regimes in our
simulation where (\ref{wphi}) differs from (\ref{wphiusual}). To
be precise, we have treated ordinary matter as a thermal
background, and have not accounted for energy flowing from $\phi$
into the heat bath. This is generally a negligible effect, except
when the dark energy density is large and $\phi$ is of
order $M_P$, which can occur at late times.
The evolution of $\phi$ can strongly influence the
thermal matter (for example, changing the coefficent of its
kinetic term), which we have not accounted for. The late time
behavior of our simulations at large $\beta$ is therefore only
qualitatively and not quantitatively correct. Observational probes
such as WMAP are sensitive to the way the dark energy density
redshifts, and hence constrain (\ref{wphi}).

\begin{center}
{\scalebox{1}{\includegraphics{plot_wosc.epsi}}}
\end{center}
\small
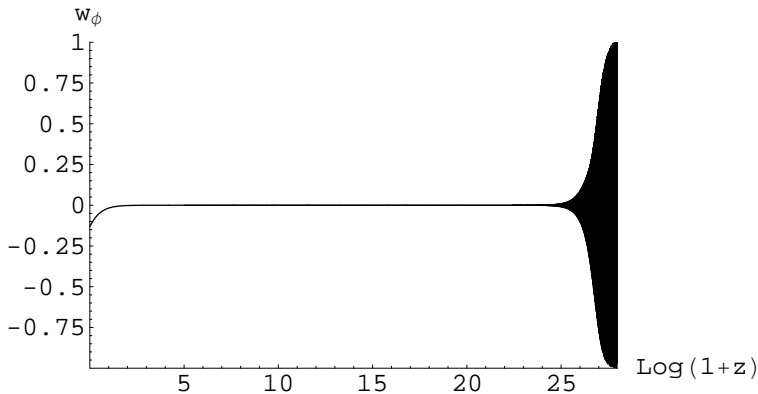
FIG. 4. The evolution of $w_{\phi}$ is shown for
$\beta= 10^2$, $l=6$
and an intermediate $\rho_{\phi i}$
($\phi_i \ll \phi_{* i}$). Note that the (very rapid)
oscillations in $w_{\phi}$
(corresponding to oscillations of $\phi$ about $\phi_*$)
are completely damped out very early in the evolution of
$\phi$, and that $w_{\phi} \sim 0$ for most of the evolution
of $\phi$.
Also, note that today $w_{\phi} \sim -0.13$, compared
to the $\beta=0$ value of $w_{\phi} \sim -0.4$.
\normalsize

\begin{center}
{\scalebox{1}{\includegraphics{plot_w0_mod.epsi}}}
\end{center}
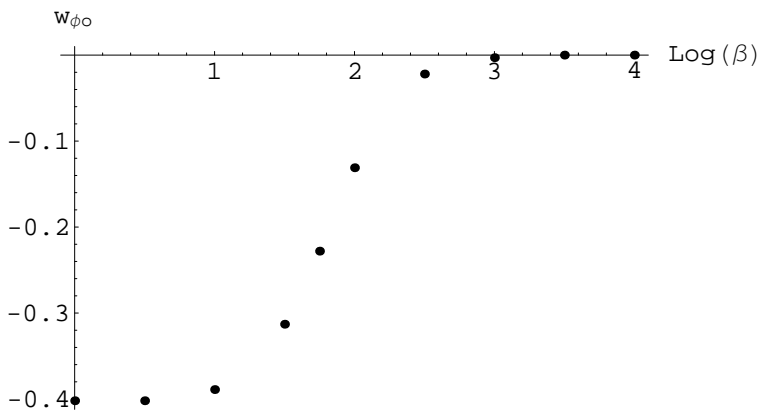
\small FIG. 5. The present value of $w_{\phi o}$ is plotted versus
$\log(\beta)$ for $l=6$. Note that for $\beta \gsim 10$, $w_{\phi
o}$ begins to deviate from its $\beta = 0$ value. For $\beta \gsim
10^3$, $w_{\phi o} \sim 0$. \normalsize

For $\beta \gsim \beta_c$, there are oscillations in
$w_{\phi}$, corresponding to the
oscillations seen in $\phi$ (see Figure 4). Again, the period of
these oscillations decreases as $\beta$ is increased, and the
period decreases exponentially with the scale factor. By
increasing $l$, the present value of $w_{\phi}$ is driven toward
zero (from below). By increasing $\beta$ (beyond $\sim
\beta_c$), $w_{\phi o}$ is also driven towards zero
(see Figure 5), as the evolution
of $\phi$ becomes controlled by $\phi_*$, and the energy density
by $V_*$.
In the extreme limit $\phi$ tracks $\phi_*$ closely and its energy
is almost entirely potential, rather than kinetic. Note that $-1 <
w_{\phi} < 0$ at late times for all $l$ and $\beta$.

For $\beta = 0$ and $l=6$, $w_{\phi o} \sim -0.4$. In general,
if a sum of inverse powers of $\phi$ are allowed, then
$w_{\phi}^{\rm eff} > -0.7$ \cite{tracker}. This
is the effective equation of state measured by supernovae and
microwave background experiments, which integrate
over a (potentially) varying $w_{\phi}$. This bound
represents the best case scenario for $\beta = 0$
tracker models, in that it is most consistent with observational
data for $w_{\phi}^{\rm eff}$, which put $w_{\phi}^{\rm eff} < -0.78$ (95\% CL)
\cite{observation}. The inclusion of the second term in
(\ref{potential}) with sufficiently large $\beta$
($\beta \gsim \beta_c$) results in $\phi \simeq \phi_*$ throughout
most of its evolution. For $l=6$, this yields an equation of state,
$w \sim 0$ (since $V_* \sim T^3$),
that is even further from the observational bound (see Figure 5).
At very late times, $\phi$ does not increase sufficiently rapidly
to stay near $\phi_*$ (see Figure 3).
Instead, it rejoins a tracker solution and its
equation of state reverts to one in which the energy density redshifts
more slowly than $V_*$ (see Figure 2).

\section{Discussion}

Our analysis shows that the evolution of the tracker field depends
quite sensitively on its interaction with thermal matter, even
when the strength of the couplings is as small as one would
imagine they may possibly be; i.e., Planck-suppressed. When
$\beta$ is larger than of order unity, there is a tendency for the
evolution to be controlled by that of $\phi_*$, once initial
oscillations have damped away. By a lucky coincidence, the
redshift of $V_* \sim T^{4l /(l+2)}$ is the same as that of the
tracker solution (during radiation domination), so that $\phi$ can
rejoin a tracker solution at late times. The most dangerous
possibility (which is realized when $\beta \gsim \beta_c$) is that
$\phi$ is still following $\phi_*$ at late times, in which case
its equation of state will be far from the observationally favored
$w_\phi = -1$. For general $l$, the equation of state obeyed by
$V_*$ is $w_{\phi *} = (l-6)/ 3(l+2)$, which is never consistent
with observational bounds for $l > 4$. We have checked that the
behavior described above is qualitatively similar when higher
order terms such as \beq \label{cubic} \left( \beta \over M_P
\right)^3 \phi^3 T^4 \eeq are included in the potential.

Hence, we conclude that there are stringent limits on the coupling
between the tracker field and any particles which are still
relativistic today, such as the photon or neutrinos. Such limits
cannot be avoided through fine tuning of the finite temperature
effective potential; once the zero temperature potential has been
computed the finite temperature effects are determined.
Interactions which are more than roughly two orders of magnitude
stronger than Planck-suppressed lead to a problematic equation of
state. Couplings of the tracker to heavy particles, which freeze
out at $T \sim m$, may alter the tracker evolution at early times,
but do not affect the observed dark energy equation of state and
are hence poorly constrained. The best hope of directly detecting
the quintessence field may be through its interaction with massive
particles.

Finally, although our analysis has focused on tracker models,
similar results apply for any quintessence
model in which the field is today slowly evolving in a very flat
potential. As we argued in the introduction, in {\it any} such
model the value of $\phi$ must be of order $M_P$ today, which
means that thermal terms such as (\ref{mass}) or (\ref{cubic})
will be important for sufficiently large $\beta$. Large couplings
to relic particles such as neutrinos can be ruled out as they lead
to a problematic equation of state.

\section*{Acknowledgements}
\noindent

The work of S.H. and B.M. was supported in part
under DOE contract DE-FG06-85ER40224.

\bigskip


\end{document}